\documentclass{iopjournal}

\usepackage{cite}
\usepackage{amssymb,amsthm,amsmath}
\usepackage{changes}

\begin{document}

\articletype{Paper} %	 e.g. Paper, Letter, Topical Review...

\title{Effect of the near-proton-emission threshold resonance in $^{11}$B on the branching ratio of beta-delayed proton emission from $^{11}$Be}

\author{Nguyen Le Anh$^{1,*}$\orcid{0000-0002-1198-9921} and Bui Minh Loc$^{2}$\orcid{0000-0002-2609-1751}}

\affil{$^1$Department of Physics, Ho Chi Minh City University of Education, 280 An Duong Vuong, Cho Quan Ward, Ho Chi Minh City, Vietnam}

\affil{$^2$San Diego State University, 5500 Campanile Drive, San Diego, CA 92182, USA}

\affil{$^*$Author to whom any correspondence should be addressed.}

\email{anhnl@hcmue.edu.vn}


\begin{abstract}
Beta-delayed proton emission from neutron halo nuclei $^{11}\mathrm{Be}$ represents a rare decay process. The existence of the narrow resonance near the proton-emission threshold in $^{11}\mathrm{B}$ explains its unexpectedly high probability. However, the accurate value of the branching ratio remains challenging to determine. We aim to quantify the influence of the narrow resonance near the proton emission threshold on the result of the branching ratio. We employ the Skyrme Hartree-Fock calculation within the potential model to obtain the branching ratio. We derive the single-particle potentials for the halo neutron and the emitting proton with minimal adjustment. Slight variations in the resonance position significantly impact the branching ratio, with the upper limit reaching the order of $10^{-5}$. Experimental determination of the resonance energy, particularly whether it lies below $200$~keV, is crucial for determining the value of the branching ratio.
\end{abstract}

\section{Introduction}

\label{introduction}

Exotic nuclei with unusual ratios of protons to neutrons can undergo a variety of decay modes~\cite{PfutznerRMP842012, VolyaFBS652024}. Occasionally, the decay probability for these rare processes, such as the beta-delayed proton emission, is unexpectedly high. The beta-delayed proton emission typically occurs in very proton-rich nuclei~\cite{BatchelderADNDT2020}. However, the emission of protons following beta decay is also energetically allowed for neutron-rich nuclei with neutrons bound by less than $782$~keV~\cite{Horoi2003}. The most promising candidate is $^{11}\mathrm{Be}$~\cite{BayePLB6962011}.
The $1/2^+$ ground state of $^{11}\mathrm{Be}$ is a one-neutron $s$-wave halo state~\cite{SchmittPRL1082012}, bound by only $501$~keV~\cite{KelleyNPA8802012}.

The beta-delayed proton emission of $^{11}\mathrm{Be}$ was confirmed by experiments indirectly~\cite{RiisagerEPJA562020} and directly~\cite{AyyadPRL1232019} with an unexpectedly high branching ratio. It can be explained by the decay that proceeds through a proton resonance in $^{11}\mathrm{B}$, located near the proton-emission threshold~\cite{RiisagerPLB7322014}. 
This resonance had not yet been discovered at that time. Its evidences were found later in experiments~\cite{AyyadPRL1292022, Lopez-SaavedraPRL1292022}.

In Ref.~\cite{AyyadPRL1232019}, which directly observed the proton emission in $^{11}\mathrm{Be}$ for the first time, the branching ratio is estimated as $8.6 \times 10^{-6}$. 
Theoretical calculations give smaller values~\cite{ElkamhawyPLB2021, AtkinsonPRC1052022, ElkamhawyPRC1082023}. 
Note that the small numerical values in Ref.~\cite{ElkamhawyPLB2021} were obtained using an experimental input parameter.
In addition, the new experiment in Ref.~\cite{SokolowskaPRC1102024} raised the question of the upper limit of the branching ratio. 
According to Refs.~\cite{SokolowskaPRC1102024, RiisagerEPJA562020}, this limit is determined to be $2.2 \times 10^{-6}$.

The resonance parameters are essential to determine the value of the branching ratio. The spin and parity of the resonance state are $1/2^+$.
However, the precise location of the resonance remains uncertain. Experimental measurements report resonance energies of $171$~keV~\cite{AyyadPRL1292022}, $197$~keV~\cite{AyyadPRL1232019}, and $211$~keV~\cite{Lopez-SaavedraPRL1292022}. From a theoretical perspective, accurately reproducing the resonance energy has proven to be challenging~\cite{volya2020, OkolowiczPRL1242020, AtkinsonPRC1052022}.
First, to impact the branching ratio's value, as shown in Ref.~\cite{BayePLB6962011}, the resonance must be in the narrow energy window that is less than $300$~keV above the proton-emission threshold.
Second, the alpha channel is open to all states in the region of interest as the alpha-emission threshold is $2.56$~MeV below the proton-emission threshold.
Shell model calculations in Ref.~\cite{volya2020} showed a $1/2^+$ state being over $1$~MeV above the proton emission threshold. The shell model embedded in the continuum calculation~\cite{OkolowiczPRL1242020} predicted that the resonance is at $142$~keV. In the \textit{ab initio} no-core shell model calculation~\cite{AtkinsonPRC1052022}, the resonance is at several MeV higher than $197$~keV, and consequently, a phenomenological shift of the resonance to $197$~keV was required to explain the large branching ratio. Despite the differences in the predicted resonance energies, these studies collectively highlight the importance of a low-lying $1/2^+$ state for understanding the beta-delayed proton emission of $^{11}$Be. While previous potential-model~\cite{BayePLB6962011,AyyadPRL1232019} and Halo-EFT studies \cite{ElkamhawyPRC1082023} demonstrated the sensitivity of the branching ratio to a near-threshold resonance, the present work provides a microscopic mean-field interpretation of the resonance and quantifies the correlation between the resonance energy and the branching ratio within a self-consistent Skyrme Hartree-Fock framework.

The analysis in Refs.~\cite{AyyadPRL1232019, OkolowiczPRL1242020, AyyadPRL1292022, Lopez-SaavedraPRL1292022} showed a large spectroscopic factor for the proton channel, suggesting that the nucleon mean-field approach provides an appropriate framework for the analysis.
Therefore, in Ref.~\cite{anh2022Be10}, using the Skyrme Hartree-Fock calculation in the continuum~\cite{dover1971, dover1972}, we focused on the single-particle properties of the resonance to analyze the low-energy proton elastic scattering data of Ref.~\cite{AyyadPRL1292022}. The results with different Skyrme forces all show the $s$-state resonance near the proton-emission threshold at $96$~keV, $152$~keV, and $376$~keV for SAMi-ISB, SkM$^*$, and SGII, respectively~\cite{anh2022Be10}. In the present work, we extend the analysis to the beta-delayed proton emission of $^{11}$Be by calculating the branching ratio within the potential model following Ref.~\cite{BayePLB6962011}, now taking into account the near proton-emission threshold resonance~\cite{AyyadPRL1292022, Lopez-SaavedraPRL1292022} which was identified as the proton $s$-state resonance~\cite{anh2022Be10}. 

The single-particle bound and scattering potentials are obtained simultaneously within the Skyrme Hartree-Fock formalism. We use two overall scaling parameters for the potentials to account for small systematic deviations of the effective interaction and to reproduce the halo neutron separation energy and the near-threshold resonance energy. Only minimal renormalization of the mean-field potential is required, indicating that both the halo structure of $^{11}\mathrm{Be}$ and the near-threshold resonance in $^{11}\mathrm{B}$ emerge naturally from the Skyrme mean field rather than being imposed by parameter tuning. The calculation method is valid for the studies of elastic scattering~\cite{dover1971, dover1972, anh2023O1415, anh2022Be10} and radiative-capture reactions at keV energy~\cite{anh2021pgamma, AnhPRC11042021, anh2022Li7}.
The flexibility of the Skyrme Hartree-Fock approach allows us to vary the near-threshold resonance energy through small changes in the proton-nucleus potential depth and thereby quantify its impact on the branching ratio.

Following Ref.~\cite{BayePLB6962011, AyyadPRL1232019}, we simplified the process to a single-channel calculation. Other channels, especially the alpha channel, are omitted~\cite{tang2025}, as including them would introduce additional uncertainties. In particular, the alpha-nucleus optical potential is poorly constrained at energies below 3~MeV, limiting the reliability of such contributions in theoretical calculations. A non-zero alpha width would contribute to the total resonance width and modify the proton elastic-scattering line shape. Nevertheless, the branching ratio is expected to remain considerably more sensitive to the resonance energy than to moderate variations of the resonance width. Moreover, when interpreting the beta-delayed proton emission of $^{11}\mathrm{Be}$ as a quasi-free neutron decay, it is particularly well-suited to our calculation, which focuses on the single channel.

The purpose of the present work is not to predict the exact resonance energy, but rather to quantify how the branching ratio depends on the experimentally constrained resonance location within a self-consistent mean-field framework. We demonstrate that the branching ratio exhibits a rapid variation when the resonance position shifts by just a few tens of keV in the energy window from $0$~keV to $281$~keV. The estimated value $8.6 \times 10^{-6}$ from Ref.~\cite{AyyadPRL1232019} is consistent to the resonance location below $200$~keV ($197$~keV~\cite{AyyadPRL1232019}). The limit of $2.2 \times 10^{-6}$~\cite{ RiisagerEPJA562020, SokolowskaPRC1102024} places the resonance at the energy above $200$~keV ($217$~keV in our calculation). The upper limit of the branching ratio may reach the order of $10^{-5}$.

From our analysis, the branching ratio is $8.98  \times 10^{-6}$. After calibration to the same resonance energy, the calculated branching ratio shows only a weak dependence on the choice of Skyrme interaction. This value is obtained for a resonance energy of $182 \pm 11$~keV, corresponding to the weighted average of the currently available experimental determinations of the near-threshold resonance energy~\cite{AyyadPRL1232019,AyyadPRL1292022,Lopez-SaavedraPRL1292022,tang2025}. The calculation adopts a spectroscopic factor of $0.51 \pm 0.06$ for the neutron halo state~\cite{LeePRC752007}. The larger values of spectroscopic factor~\cite{SchmittPRL1082012} increase the branching ratio even to the order of $10^{-5}$. An experiment measuring the resonance energy to ascertain whether it falls below $200$~keV will be key to resolving the puzzle.

\section{Method of calculation}
\label{method}

The beta-delayed proton emission from $^{11}\mathrm{Be}$ is
\begin{equation}
    ^{11}\text{Be} \to {}^{10}\text{Be} + p + e^- + \bar{\nu}_e.
\end{equation}
The branching ratio ($b_{\beta p}$) is given by
\begin{equation}
    b_{\beta p} = \dfrac{W T_{1/2}}{\ln 2},
\end{equation}
where $T_{1/2} = 13.76 \pm 0.07$ s~\cite{audi2003} is the half-life and $W$ is the decay probability per unit time of the halo nucleus $^{11}\mathrm{Be}$, defined as
\begin{equation}
    W = \int_0^Q \dfrac{dW}{dE}\,dE ,
\end{equation}
where $E$ is the relative energy of the proton-core system in the exit channel, and $Q$ is the total energy available in the decay, expressed as
\begin{equation}
    Q = (m_n - m_p - m_e)c^2 - S_n,
\end{equation}
where $m_n$, $m_p$, and $m_e$ represent the neutron, proton, and electron masses, respectively, and $S_n$ is the neutron separation energy of the halo nucleus. In this case, $S_n = 501$~keV and $Q = 281$~keV.

The $ft$ value for the contribution of both Fermi and Gamow-Teller transitions is given by
\begin{equation}
    ft = \dfrac{K}{B_{\text{F}}+\lambda^2 B_{\text{GT}}},
\end{equation}
where $K = 6144$ s~\cite{towner2010,PfutznerRMP842012}.
The differential decay probability per unit time is given by~\cite{BayePRC822010, BayePLB6962011}
\begin{equation}
    \dfrac{dW}{dE} = \dfrac{1}{2\pi^3} \dfrac{m_e c^2}{\hbar} G_\beta^2 f(Q-E) \left( \dfrac{dB_{\text{F}}}{dE} + \lambda^2 \dfrac{dB_{\text{GT}}}{dE} \right),
    \label{eq:dW/dE}
\end{equation}
where $G_\beta = 2.996 \times 10^{-12}$ is the dimensionless beta-decay constant, and $\lambda = -1.268$ is the axial-to-vector coupling constant ratio.
In Eq.~\eqref{eq:dW/dE}, the Fermi integral $f(Q-E)$ is evaluated with respect to $w_e$, which is the total energy of the electron (in units of $m_ec^2$) at a given phase-space point and is given by
\begin{equation} \label{eq:integral_Z}
    f(w_0) = \int_1^{w_0} p_e w_e (w_0 - w_e)^2 F(Z,w_e)\,dw_e,
\end{equation}
where $p_e$ is the electron momentum, defined as $p_e = \sqrt{w_e^2 - 1}$.
The total energy of the electron, including its rest mass, is $w_0 = 1 + (Q - E)/(m_e c^2)$. The Fermi function $F(Z,w_e)$ accounts for the Coulomb interaction between the emitted electron and the charge of the daughter nucleus ($Z = 4$). For low-energy beta decays, the Fermi function can be approximated by
\begin{equation} \label{eq:FZE}
    F(Z,w_e) \approx \dfrac{2\pi\eta_e}{1-\exp(-2\pi\eta_e)},
\end{equation}
where the Sommerfeld parameter $\eta_e$ is defined as $\eta_e = \alpha Z w_e/p_e,$ with $\alpha=1/137$ being the fine-structure constant. 

The initial state $i$ is a bound state where a neutron couples to the core nucleus ${}_Z^A X$. The final state $f$ is a scattering state involving the core and a proton. The orbital angular momenta of the relative motion in the initial and final states are denoted as $\ell_i$ and $\ell_f$, respectively, while the total angular momenta are  $j_i$ and $j_f$.

The reduced decay probabilities for Fermi ($B_{\text{F}}$) and Gamow-Teller ($B_{\text{GT}}$) transitions are given by~\cite{BayePRC822010, BayePLB6962011}
\begin{equation} \label{eq:dBF-GT}
    \dfrac{dB_{\text{F}}}{dE} = \dfrac{1}{\hbar v} I_{if}^2(E), \quad \text{and} \quad \dfrac{dB_{\text{GT}}}{dE} = 3 \dfrac{dB_{\text{F}}}{dE},
\end{equation}
where $E$ and $v$ are the relative energy and velocity of the system in the scattering state, respectively, $E = \mu v^2/2$, and $\mu$ is the reduced mass. As the final wave function is independent of $j_f$, the Gamow-Teller term simplifies to three times the Fermi term.

The key component is the overlap radial integral given by
\begin{equation}
    I_{if}(E) = S_F^{1/2} \int_0^\infty \chi(E,r) \phi(r) \,dr,
    \label{theOverlap}
\end{equation}
where $S_F$ is the neutron-halo spectroscopic factor associated with the initial-state configuration $^{11}\mathrm{Be}\approx{}^{10}\mathrm{Be}+n$. 
The similar radial integral in Eq.~\eqref{theOverlap} is successfully applied to study the magnetic dipole transition ($M1$) in light nuclei at keV energy~\cite{anh2022Li7}. The $s$-wave functions $\chi(E,r)$ and $\phi(r)$ are obtained by solving the Schr\"odinger equations of the initial (bound) and final (scattering) states of neutrons ($q=0$) and protons ($q=1$), respectively. The two potentials are obtained within the Skyrme Hartree-Fock formalism
\begin{equation} \label{pot}
V_q (E,r) = \dfrac{m_q^*(r)}{m} \left\{V_q^{\rm HF}(r) + \dfrac{1}{2} \dfrac{d^2}{dr^2}\left(\dfrac{\hbar^2}{2m_q^*(r)} \right)  - \dfrac{m_q^*(r)}{2\hbar^2} \left[\dfrac{d}{dr}\left(\dfrac{\hbar^2}{2m_q^*(r)}\right)\right]^2 \right\}
+ \left[1 - \dfrac{m_q^*(r)}{m}\right] E.
\end{equation}
The Hartree-Fock mean field $V_q^{\rm HF}(r)$ and the effective mass $m_q^*(r)$ \ are obtained from the \texttt{skyrme\_rpa} program~\cite{ColoCPC1842013}. The energy $E$ can be either positive for the scattering state or negative for the bound state in Eq.~\eqref{pot}. We use the Skyrme forces SkM*~\cite{skm_bartel1982}, SAMi-ISB~\cite{SAMiISB_maza2018}, SLy4~\cite{SLy4_chabanat1998}, and SAMi~\cite{SAMi_maza2012} for illustration.

When using Eq.~\eqref{pot}, the two nuclear potentials are adjusted by two overall scaling parameters $\mathcal{N}_i$ and $\mathcal{N}_f$ for the initial and final states, respectively. The parameter $\mathcal{N}_i$ is fine-tuned to reproduce the separation energy of the halo neutron, $S_n = 501$~keV, and the parameter $\mathcal{N}_f$ is calibrated to the location of the resonance. Note that the overlap function in the study is different from the one in the radiative capture reaction by the fact that the initial and final states are different particles. The many-body Coulomb potential is included for the proton case and kept unchanged. For the elastic scattering, the cross section is obtained by the standard matching method, and the width $\Gamma_p$ is
\begin{equation}
    \Gamma_p(E_R) = 2 \left[\frac{d\delta_\ell(E)}{dE} \right]^{-1}_{E=E_R},
\end{equation}
where $\delta_\ell$ is the phase shift and $E_R$ is the resonance location. 

Finally, the spectroscopic factor $S_F$ in Eq.~\eqref{theOverlap} is the percentage of the halo-neutron configuration in $^{11}\mathrm{Be}$. It leads to a reduction in the branching ratio and an enhancement in the $ft$ value. The spectroscopic factor $S_F$ is $ 0.71$ according to Ref.~\cite{SchmittPRL1082012}. Recently, Ref.~\cite{tang2025} reported the extracted value of $S_F = 0.25 \pm 0.06$. We use the adopted value of $S_F = 0.51 \pm 0.06$ provided in Ref.~\cite{LeePRC752007} for all calculations. It is noted that no additional spectroscopic factor is introduced for the final $^{11}$B resonance, which is described directly by the continuum Skyrme Hartree-Fock scattering wave function. The values for parameters and constants in the classical theory of the weak interaction, such as $K$ and $\lambda$, contain uncertainties. In the present work, we emphasize the uncertainty caused by the location of the nuclear single-particle resonance, which can lead to order-of-magnitude variations in the branching ratio.

\section{Results and discussion}
%%\label{}

\begin{figure}[]
    \centering
    \includegraphics[width=0.6\linewidth]{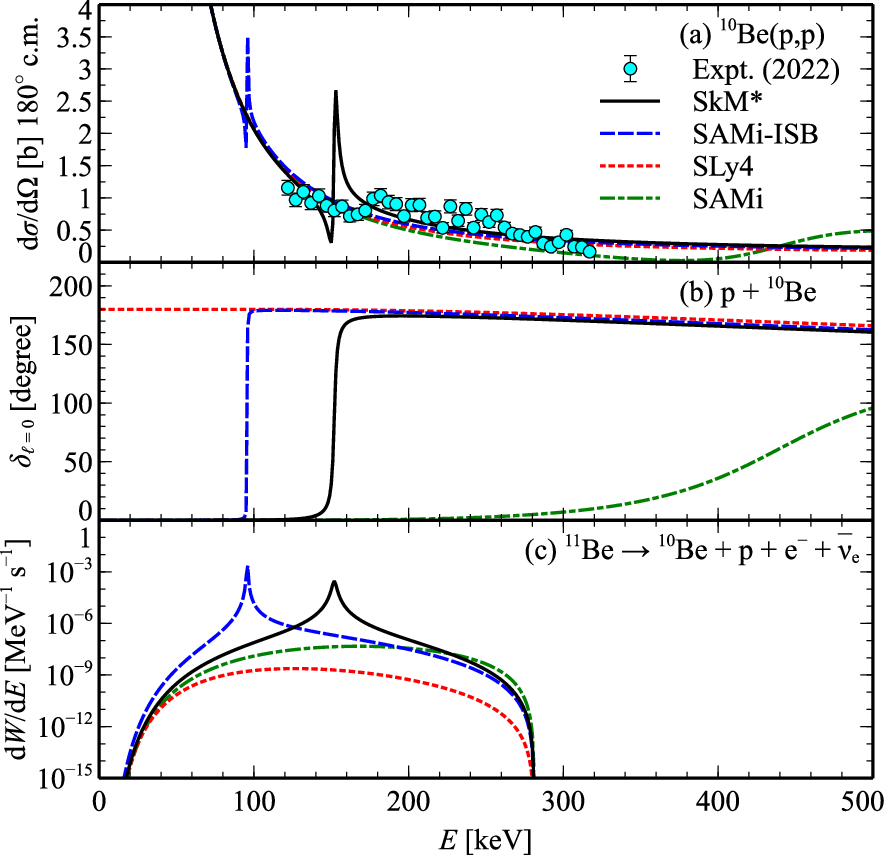}
    \caption{(a) Excitation functions in the center-of-mass frame for elastic $^{10}$Be$(p,p)$ scattering calculated with the SkM$^*$, SAMi-ISB, SLy4, and SAMi interactions before calibration ($\mathcal{N}_f=1$). The experimental data are taken from Ref.~\cite{AyyadPRL1292022}. (b) Corresponding $s$-wave phase shifts for the $p+{}^{10}$Be system. (c) Decay-probability distributions per unit energy for the beta-delayed proton emission of $^{11}$Be calculated with the same interactions.}
    \label{fig:untuned}
\end{figure}

Before any calibration, the Skyrme Hartree-Fock calculations already predict qualitatively different scenarios for the $s_{1/2}$ proton state in $^{11}\mathrm{B}$, as shown in Figure~\ref{fig:untuned}. 
For the SkM$^*$ and SAMi-ISB interactions, resonances appear at $152$~keV and $96$~keV, respectively, while the SLy4 interaction predicts a bound state below the ${}^{10}\mathrm{Be}+p$ threshold, and the SAMi interaction places the resonance at $511$~keV, above the total decay energy $Q$. 
This diversity reflects the well-known difficulty of describing near-threshold states within mean-field approaches, but it also reveals that the emergence of an $s$-wave state close to the proton-emission threshold is already encoded in the underlying nuclear mean field.

\begin{table}[b]
    \centering
    \caption{Branching ratios obtained with and without calibrating to the exact resonance locations at $171$~keV, $182$~keV, and $211$~keV. The spectroscopic factor $S_F$ of the halo neutron state in $^{11}\mathrm{Be}$ is $0.51$. The calculations with $\mathcal{N}_f = 1$ are noted as ``Untuned''. When the factor $\mathcal{N}_f$ is adjusted, all Skyrme forces give slightly different results. The upper limit of the branching ratio may reach the order of $10^{-5}$.
    }
    \label{tab:br}
    \begin{tabular}{cccccccc}\\
    \hline \hline 
        Resonance & SkM$^*$ & SAMi-ISB & SLy4 & SAMi  \\  \hline 
        Untuned & $23.2\times 10^{-6}$ & $37.0 \times 10^{-6}$ & $2.26 \times 10^{-9}$ & $4.89 \times 10^{-8}$  \\
        $171$~keV & $13.2\times 10^{-6}$ & $12.9 \times 10^{-6}$ & $12.9 \times 10^{-6}$ & $12.9 \times 10^{-6}$ \\
        $182$~keV & $8.91\times 10^{-6}$ & $8.98 \times 10^{-6}$ & $8.97 \times 10^{-6}$ & $9.05 \times 10^{-6}$ \\
        $211$~keV & $2.86\times 10^{-6}$ & $2.90 \times 10^{-6}$ & $2.88 \times 10^{-6}$ & $2.89 \times 10^{-6}$ \\
        \hline \hline 
    \end{tabular}
\end{table}

A crucial observation is that only when the resonance lies below $281$~keV does the branching ratio increase dramatically. 
The presence of such a near-threshold $s$-wave resonance enhances the branching ratio by up to three orders of magnitude compared to cases without a resonance in this energy window. 
The corresponding branching ratios obtained without any calibration ($\mathcal{N}_f = 1$) are listed in Table~\ref{tab:br} and labeled ``Untuned''.

\begin{table}[b]
\caption{Parameters $\mathcal{N}_i$ and $\mathcal{N}_f$ used to reproduce the neutron separation energy $S_n$ in $^{11}\mathrm{B}$, and the resonance location at $182$~keV, respectively.}
    \label{tab:lambda}
    \centering
    \begin{tabular}{cccccccc}\\
    \hline \hline
        Parameter & SkM$^*$ & SAMi-ISB & SLy4 & SAMi \\ \hline
        $\mathcal{N}_i$ & $1.086$ & $1.141$ & $1.149$ & $1.125$  \\
        $\mathcal{N}_f$ & $0.997$ & $0.991$ & $0.978$ & $1.028$ \\
        \hline \hline
    \end{tabular}
\end{table}

The calibration procedure is not intended to generate the near-threshold $s_{1/2}$ state. Rather, it introduces only a small adjustment of the self-consistent mean-field potential to constrain the resonance energy to the experimentally observed region, thereby enabling a systematic investigation of the correlation between the resonance location and the branching ratio.
When the calibration is applied to fix the location of the resonance, the results remain nearly identical with all Skyrme forces in the study. 

The values of $\mathcal{N}_i$ and $\mathcal{N}_f$ as shown in Table~\ref{tab:lambda} are close to unity. For the neutron $2s_{1/2}$ state, the Hartree-Fock single-particle energies are 496, 202, 405, and 446 keV for the SLy4, SkM$^*$, SAMi, and SAMi-ISB interactions, respectively. These values indicate that the state already lies very close to the threshold. All $\mathcal{N}_i$ are slightly larger than unity, showing that the energy of the $s_{1/2}$ state without the fine-tuning is positive and very close to zero. A minimal adjustment can make this state bound and reproduce the neutron-halo properties of $^{11}\mathrm{Be}$. This demonstrates that the near-threshold character of the $s$-wave state is not imposed by construction but emerges naturally from the self-consistent mean-field framework. Furthermore, the weak dependence of the branching ratio on the choice of Skyrme force demonstrates that the correlation between the resonance position and the branching ratio is robust and largely interaction-independent. Therefore, the strong sensitivity of the beta-delayed proton branching ratio to the resonance energy reflects a structural mechanism associated with near-threshold single-particle states rather than a model-dependent artifact. The sensitivity of the resonance position, particularly in the narrow energy window that strongly and directly affects the branching ratio, to the value of $\mathcal{N}_f$ was discussed in Ref.~\cite{anh2022Be10}. 

\begin{figure}[]
    \centering
    \includegraphics[width=0.6\linewidth]{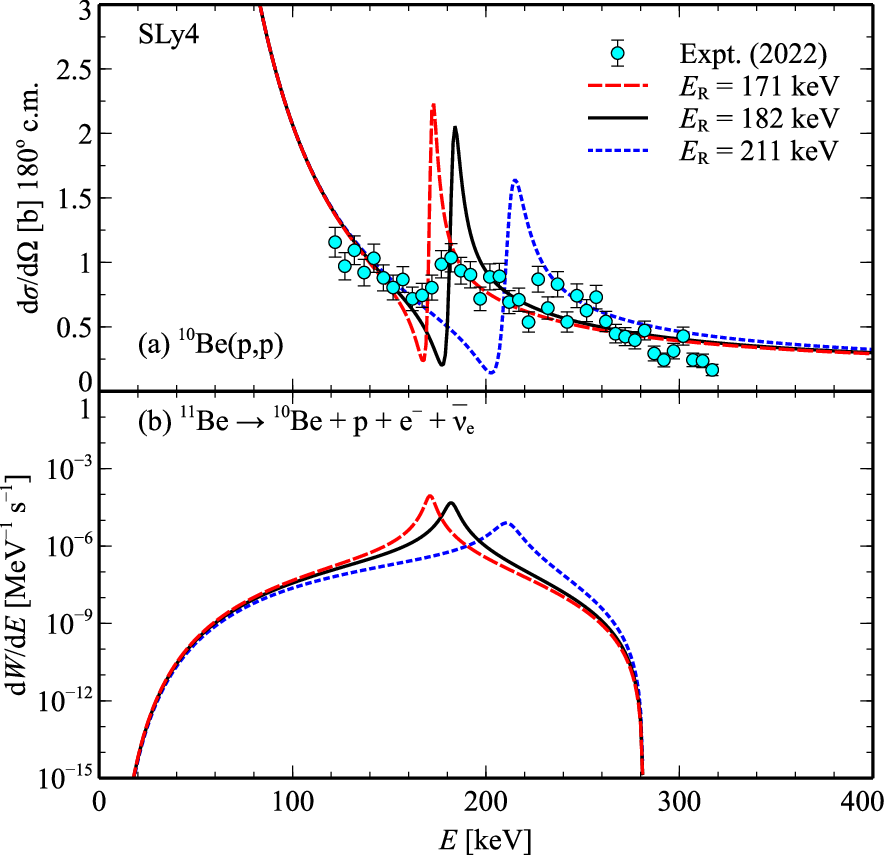}    \caption{(a) The calculated excitation functions in the center-of-mass frame for $^{10}\mathrm{Be}(p,p)$ compared with the experimental data from Ref.~\cite{AyyadPRL1292022}. (b) The decay probability distribution per second for SLy4 fitted to the exact resonance locations at $171$~keV~\cite{AyyadPRL1292022}, $211$~keV~\cite{Lopez-SaavedraPRL1292022}, and the average value of $182$ keV.}
    \label{fig:VaryRes}
\end{figure}

The evidence of the near-threshold resonance in Refs.~\cite{AyyadPRL1292022, Lopez-SaavedraPRL1292022} confirmed the mechanism of the enhanced branching ratio.
However, the experimental uncertainties in these studies are significant, with error bars being $\pm 20$~keV~\cite{AyyadPRL1292022} and $\pm 40$~keV~\cite{Lopez-SaavedraPRL1292022}. As shown in our analysis in Table~\ref{tab:br} and illustrated in Figure~\ref{fig:VaryRes}, the value of the branching ratio is strongly sensitive to the location of the resonance.

\begin{table}[]
    \centering
    \caption{The overlap integrals $I_{if}$, $B_{\text{F}}$ and $B_{\text{GT}}$ transitions, $\log(ft)$ values, and the widths of single-particle resonances $\Gamma_p$ computed with SLy4 force.}
    \label{tab:ERGamma}
    \begin{tabular}{lccc}\\
    \hline \hline 
        $E_R$ [keV] & 171 & 182 & 211  \\  \hline 
        $\Gamma_p$ [keV] & $4.738$ & $6.249$ & $11.905$ \\
        $I_{if}$ & $19.27$ & $17.02$ & $12.82$ \\
        $B_{\text{F}}$ & $1.364$ & $1.353$ & $1.307$ \\
        $B_{\text{GT}}$ & $4.092$ & $4.061$ & $3.920$ \\ 
        $\log(ft)$ & $2.621$ & $2.628$ & $2.659$ \\ 
        \hline \hline 
    \end{tabular}
\end{table}

\begin{figure}[]
    \centering
    \includegraphics[width=0.6\linewidth]{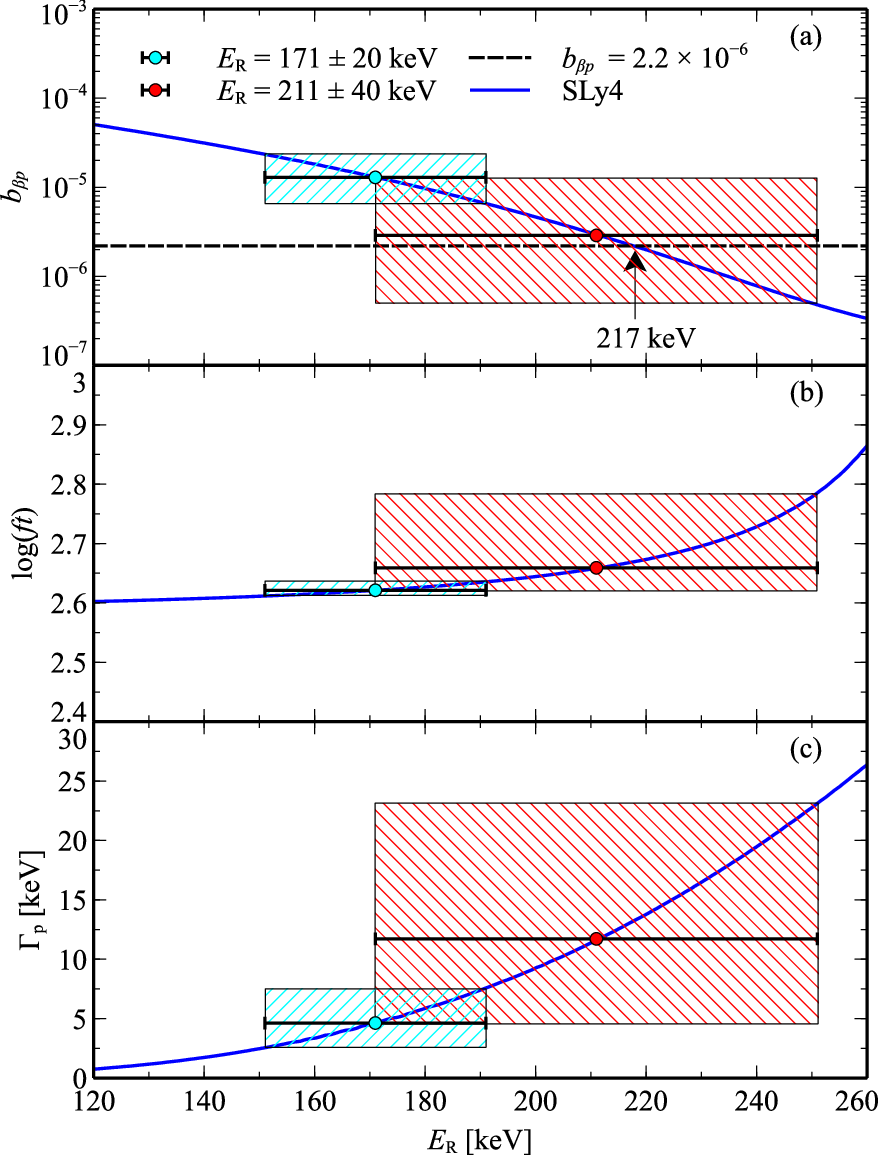}
    \caption{The branching ratios $b_{\beta p}$, logarithm of $ft$ values, and proton-decay width $\Gamma_p$ as functions of resonance locations $E_R$ with the error bars. The value of $2.2 \times 10^{-6}$ reported in Refs.~\cite{ RiisagerEPJA562020, SokolowskaPRC1102024} places the resonance at $217$~keV.}
    \label{fig:blogftGamma}
\end{figure}

Note that the branching ratio in Ref.~\cite{AyyadPRL1232019} is $8.6 \times 10^{-6}$ based on the overlap $I = 53$ and an $S_F = 0.34$. In Table~\ref{tab:ERGamma}, we show the value of the overlap $I$ and the width of the resonance from our calculations. The values of $B_{\text{F}}$ and $B_{\text{GT}}$ are also given.
The branching ratio is in the order of $10^{-5}$ if the resonance is at $171$~keV. The smaller value of the branching ratio corresponds to the resonance above $200$~keV. For example, if the resonance is at $217$~keV, the branching ratio is $2.2 \times 10^{-6}$ as shown in Figure~\ref{fig:blogftGamma}. Finally, our analysis gives the value of the branching ratio as $8.98 \times 10^{-6}$ using $S_F = 0.51$ in Ref.~\cite{LeePRC752007}. A higher value $S_F = 0.71$ given in Ref.~\cite{SchmittPRL1082012} increases the value of the branching ratio up to $12.78 \times 10^{-6}$, i.e., at the order of $10^{-5}$. 

Our present study employs only the single-particle mean-field approach. Notably, the ${}^7\mathrm{Li}+ \alpha$ channel has a threshold of $2.56$~MeV below the proton-emission threshold of $^{11}\mathrm{B}$. However, a more sophisticated treatment that includes the alpha channel faces a much greater challenge. An optical model potential for alpha scattering below $3$~MeV, even a phenomenological one, is not reliable. Adding the alpha channel may introduce more uncertainty to the problem, especially since the result is extremely sensitive to the input. Recent measurements have also determined the partial alpha-decay widths of this resonance, providing additional constraints on the competition between the proton-decay and alpha-decay channels~\cite{tang2025}. The small alpha spectroscopic factor extracted from the experiment indicates that the state has a predominantly single-particle $^{10}\mathrm{Be} + p$ character rather than an alpha-cluster configuration. As shown in Table~\ref{tab:ERGamma}, the calculated width $\Gamma_p$ increases when the location of the resonance $E_R$ increases. 

\begin{figure}[]
    \centering
    \includegraphics[width=0.6\linewidth]{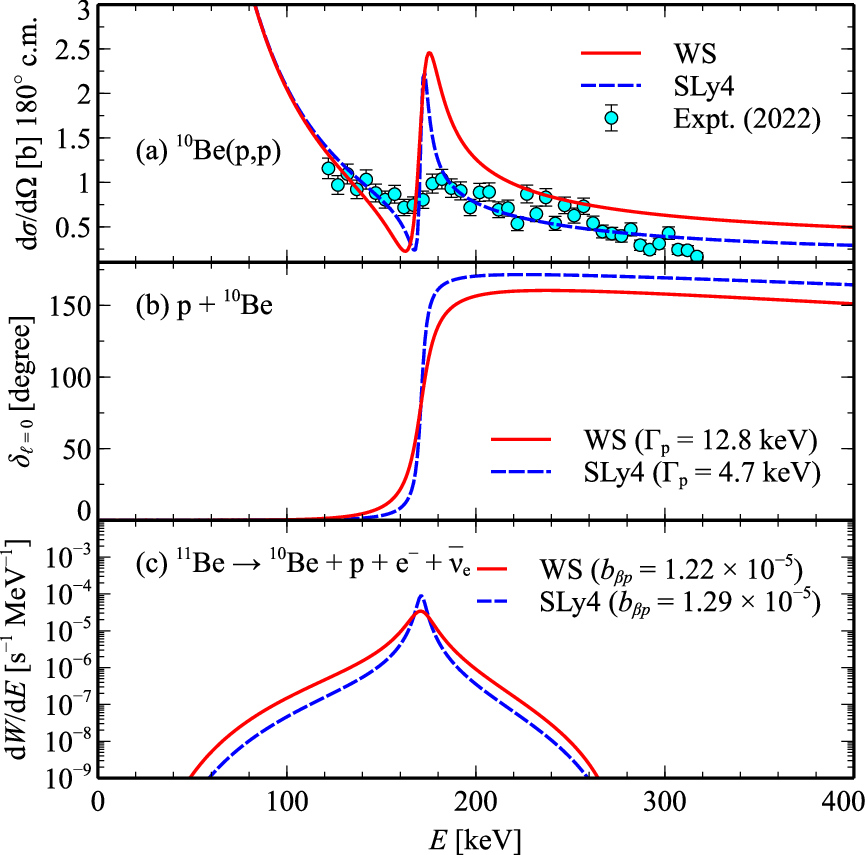}
    \caption{Same as Figure~\ref{fig:untuned} but for the comparison of two single-channel $^{10}\mathrm{Be}+p$ calculations with nearly identical resonance energies but different resonance widths, illustrating the weak dependence of the branching ratio on the resonance width.}
    \label{fig:WS_br}
\end{figure}

To further examine the role of the resonance width, we performed additional single-channel calculations using Woods-Saxon potentials fitted to reproduce nearly the same resonance energy but different resonance widths. A broader resonance with $\Gamma_p=12.8$ keV was obtained by modifying the Woods-Saxon geometry and depth parameters relative to the Hartree-Fock-based parametrization. The results are shown in Figure~\ref{fig:WS_br}. Although the resonance width increases from $4.7$ keV to $12.8$ keV, the calculated branching ratio changes only from $1.29\times10^{-5}$ to $1.22\times10^{-5}$. This result suggests that the branching ratio is considerably more sensitive to the resonance energy than to moderate variations of the resonance width. A quantitative treatment of the open ${}^7\mathrm{Li}+\alpha$ channel would require coupled-channel calculations and reliable alpha-nucleus optical potentials at very low energies, which are beyond the scope of the present work.

\section{Conclusion}
%%\label{}

Our result strengthens the link between low-energy proton elastic scattering, a nuclear process governed by strong interactions, and the branching ratio of beta-delayed proton emission, a weak interaction process. This connection highlights the interplay between nuclear structure and weak decay channels, offering deeper insights into how nuclear and weak forces influence each other at low energy. Our results provide a microscopic interpretation of why small shifts in the resonance energy, at the level of tens of keV, can induce order-of-magnitude variations in weak-decay observables. The results provide a quantitative benchmark for ongoing and future measurements of near-threshold proton resonances, offering a direct criterion for assessing the consistency between elastic scattering data and beta-decay observables. 

As suggested in Ref.~\cite{SokolowskaPRC1102024}, further studies are needed to verify the existence of the $\beta p$ decay branch of $^{11}\mathrm{Be}$ and to clarify the puzzle of its strength. Determining the resonance location with greater precision is crucial to solving the puzzle~\cite{AyyadPRL1242020Erratum}. In addition, the near-threshold alpha elastic scattering experiment for $^{7}$Li is desired to better constrain the effect of the alpha channel.
High-profile FRIB experiments with attention to near-threshold physics are now available~\cite{brownFRIB2024}.

\ack{We would like to thank Prof. Vladimir Zelevinsky for valuable discussions. This research is funded by the Vietnam National Foundation for Science and Technology Development (NAFOSTED) under grant number 103.04-2025.07. BML is supported by the U.S. Department of Energy, under Award Number DE-NA0004075.}

%\funding{Sample text inserted for demonstration.}
% This section is a list of funder names and grant numbers

%\roles{Sample text inserted for demonstration.}
% List author names and the contributions made to the article, using terms from the NISO Contributor Roles Taxonomy (CRediT) https://credit.niso.org

\data{The data cannot be made publicly available upon publication because no suitable repository exists for hosting data in this field of study. The data that support the findings of this study are available upon reasonable request from the authors}
% For more information on IOP Publishing's research data policy see: https://publishingsupport.iopscience.iop.org/questions/research-data/

%\suppdata{Sample text inserted for demonstration.}

%\section*{References}
\bibliographystyle{iopart-num}
\bibliography{refs}

\end{document}